\documentclass[12pt]{article}
\begin{document}
 
\title{Some aspects of the orientation of galaxies in clusters
\footnote {Presented at The Sixth Scientific Conference "Selected Issues of
Astronomy and Astrophysics" in honor of Bohdan Babiy 4-6 October 2011 Lviv.}
}
\medskip
 
\author{
Paulina Pajowska ${^1}$ W{\l}odzimierz God{\l}owski${^2}$ \\
Elena Panko${^3}$ Piotr Flin $^{4}$
}
\maketitle

1. Uniwersytet Opolski, Institute of Physics, ul.  Oleska  48,
45-052 Opole, Poland e-mail:  paoletta@interia.pl

2. Uniwersytet Opolski, Institute of Physics, ul.  Oleska  48,
45-052 Opole, Poland e-mail:  godlowski@uni.opole.pl

3. Kalinenkov Astronomical Observatory, Nikolaev State
University, Nikolaev, Ukraine email: panko.elena@gmail.com 
 
4. Pedagogical University, Institute of Physics, 25-406 Kielce, ul.
Swietokrzyska 15, Poland  e-mail: sfflin@cyf-kr.edu.pl

\section*{Abstract}
\medskip
The analysis of Tully's groups of galaxies belonging to the Local Supercluster
(LSC) was performed. In the 1975 Hawley and Peebles \cite{h4} presented the method
for investigations of the galaxies orientation in the large structures.
In our previous papers \cite{g10a,g11a} statistical test  proposed by Hawley
and Peebles for investigation of this problem was analyzed in details and some
improvements were suggested. On this base the new method of the analysis of
galactic alignment in clusters was proposed. Using this method, God{\l}owski
\cite{g11b} analyzed the orientation of galaxies inside Tully's group founding
no significant deviations from isotropy both in orientation of position angles
and $\delta_D$ and $\eta$ angles as well, giving the spatial orientation of galaxy
planes. In the present paper we examined carefully and methodically the dependence of alignment in Tully's
groups on morphological type of galaxies. Moreover, we discussed the consequences
of different approximation of "true shape" of the galaxies for different
morphological types, possible influence of this problem for investigation of
spatial orientation of galaxies. In addition, we discussed the implications of the obtained results for
the theory of galaxy formation as well.
 
{\bf keywords}
 angular momenta, galaxies, PACS 98.65.-r, 98.62.Ai
 
\section{Introduction}
The aspect of structure formation in the Universe is one of the crucial
problem of the modern extragalactic astronomy and cosmology. Moreover,
since different scenarious of structure formation predicts different
orientations of galxies belonging to that structures
\cite{Peebles69,Zeldovich70,Sunyaew72,Doroshkevich73,Shandarin74,Wesson82,Silk83,Dekel85,Bower05},
the investigation of the orientations of galaxies planes is regarded as a
standard test of galaxies formation scenarios.
 
An interesting aspect of this problem consists in analysing the orientation of
galaxies inside the galaxy structures. For the latest review of this problem
see \cite{g11}. The very important question is whether exists the
dependence on the alignment to the mass of the analyzed structure or not.
God{\l}owski et al. \cite{g05} suggested that alignment of galaxies in cluster 
should increase with the number of a particular objects in the individual cluster.
That hypothesis was confirmed qualitatively by Aryal et al. \cite{Aryal07}.
God{\l}owski at al. \cite{g10a} verified this suggestion analysing sample
of 247 rich Abell clusters using statistical tests and found out that
alignment increases with the richness of the clusters.
 
Another aspect of this problem concerns with the orientation of galaxies in 
less massive, poor galaxy structures - groups of galaxies. It also should be noticed 
that for groups and clusters of galaxies there
is no evidence of rotation. Moreover, Hwang and Lee \cite{Hwang07}  examined dispersions and
velocity gradient of 899 Abell clusters and found a possible rotation in only six of
them. Thus, any non-zero angular momentum of groups and clusters of galaxies would
just come from possible alignment of galaxy spins. 

The orientation of galaxies in clusters was investigated many times.
Thompson \cite{Tom76} found alignment of galaxy orientations in the Virgo and A2197 clusters.
Adams \cite{Ada80}  discovered a bimodal distribution of galaxy orientations by examining
the combined data for seven galaxy clusters (A76, A179, A194, A195, A999, A1016,
A2197). The orientation of principal axes of the clusters corresponded with one of
those maxima. Helou and Salpeter \cite{Hel82}, studying 20 galaxies belonging to the Virgo
cluster, found that their spins are not directed in random, however the nature of
this nonrandom distribution was not too clear. MacGillivray and Dodd \cite{MacG85a} investigated
the distribution of orientation of galaxies in the Virgo cluster and showed
that the galaxy planes are perpendicular to "the direction towards the cluster's
center", i.e. the galaxies' rotational axes are aligned towards that center. 

On the other hand, Bukhari \cite{Buk88} as well as Bukhari and Cram \cite{Buk03} studying 
orientation of galaxies within clusters, did not recognize any alignment. Han et al. \cite{Han95}
probed a region of the LSC with an enhanced density galaxies. They analyzed a sample of 
60 galaxies with well-known spins  founding no alignment. Flin and Olowin \cite{Flin91} 
Trevese et al. \cite{Tre92} and Kim \cite{Kim01} investigating isolated Abell clusters,
detected just rudimentary traces of alignment. The similar results were obtained
by Torlina et al. \cite{Tor07} from studies of the Coma cluster and its vicinity. 
Gonzalez and Teodoro \cite{Gon10} interpreted the alignment of just the brightest galaxies 
within a cluster as an effect of action of gravitational tidal forces. 

Summing up the results obtained by various authors, it can be stated that we
have no satisfactory evidence to support the galaxy axis alignment in the groups
and poor clusters of galaxies, while there is ample evidence of this kind for the rich
clusters of galaxies. Additionally, it is obvious that in the isolated Abell groups the 
brightest galaxies manifest a rudimentary alignment,\cite{Flin91,Tre92,Kim01}  while in 
the most numerous clusters a non-random galaxy orientation alignment was
found \cite{Aryal07,g4,Baier03,Djo83,Kit03,Wu98}.

The alignment of galaxies in the Tully's groups of galaxies \cite{t3}
was analyzed for the first time by God{\l}owski and Ostrowski \cite{g5}.
For each cluster they studied the $\Delta_{11}$ parameter
describing  the galactic axes alignment with respect  to  a chosen
cluster pole, divided by its formal error $\sigma(\Delta_{11})$ ($s
\equiv \Delta_{11} / \sigma(\Delta_{11})$). The cluster pole coordinates
change  along the entire  celestial sphere.  The  resulting maps were
analyzed  for correlations  of their maxima with the important points on
the maps. It was found that maxima correlate well with the  direction of the  
line of sight. God{\l}owski and Ostrowski \cite{g5}  concluded that this strong 
and systematic effect, was generated by the process of galactic axis de-projection
from its optical image, is present in the catalogue data.
 
Tully's groups were investigated again by God{\l}owski et al. \cite{g05}.
In that paper it was found that the groups do not exhibit a clear
evidence for existence of the alignment in the investigated structure. However,
they concluded that the observational effect generated by the process of
de-projection of galaxies \cite{g5} and later confirmed by \cite{g4,Baier03},
mask any possible alignment with a high degree. For that reason the more detailed
studies of the orientation of galaxy in Tully's groups of galaxies were required.
In the paper by God{\l}owski \cite{g11b} it was shown that using "true shape" i.e.
true axis ratio of galaxies $q_0$ depending on morphological type according to 
Heidmann et al.\cite{hhv} [hereafter HHV] with help of Fouque \& Paturel \cite{fp85} 
[hereafter FP] corrections of $q$ to standard photometrical axial ratios, allowed us 
to avoid this problem. This gives much more powerful investigations of the spatial 
orientation of galaxies.
 
Moreover, using new method of the analysis of the galactic alignment in clusters
proposed by \cite{g11a} on the base of statistical test proposed by Hawley and
Peebles \cite{h4}, God{\l}owski \cite{g11b}
analyzed the orientation of galaxies inside Tully's groups founding
no significant deviation from isotropy both in orientation of position angles
and $\delta_D$ and $\eta$ angles giving the spatial orientation of galaxy
planes. In the present paper we analyzed the dependence of the alignment in Tully's
groups on morphological type of galaxies.

\section{Observational data}
 
We analyzed the alignment of galaxies in galaxy groups belonging to the LSC.
Groups were taken from Tully Nearby Galaxies (NBG) listed in the Catalogue \cite{t3}.
This Catalogue contains 2367 galaxies with radial velocities less than
$3000\,km\,s^{-1}$ which give the posibility of remove the background objects.
Tully's Catalogue provides relatively uniform coverage of the entire unobscured
sky \cite{t2}. Catalog do not contain the information about position angles
of galaxies. It is the reason that position angles were taken from \cite{n1,n2,l1,l2}
while some missing measurements  were made on Palomar Sky Survey prints by Flin
\cite{g10}. The NBG Catalogue gives the group affiliation for the galaxies
belonging to the catalogue. In our opinion, the groups extracted from the NBG Catalogue
are one of the best selections with precise criterion of groups membership.
The galaxy distances are based on velocities, assuming the flat cosmological model 
($q_0 = 1/2$) with $H_0 = 75 km s^{-1} Mpc^{-1}$ and the model describing velocity
perturbations in the vicinity of the Virgo Cluster \cite{ts1} 
It means, that the  galaxy distances are very  well and in uniform maner
determined. As a result the lists of galaxies belonging to the particular
groups are free from the foreground and background objects which is crucial
in such type of the analysis. For our study we use only those groups from NBC 
Catalogue which have at least 40 members.

\section{Methods of the investigations}
 
One should note that two main methods for the study of the
galaxy orientation were proposed till now. In the first one \cite{h4}
the distribution of the position angle of the galactic image major
axis was analyzed. The second approach is based on the de-projection
of the galaxy images, where the  galaxy's inclination with respect
to the observer's line of sight $i$ is  considered. The latter method 
allowed us to use also the face-on galaxies, in contrast to investigations 
of the position angles where face-on and nearly face-on galaxies should be 
excluded from the consideration.
This method was originally proposed by {\"O}pik \cite{Op70},
applied by Jaaniste \& Sarr \cite{Ja78} and significantly modified by
Flin \& God{\l}owski \cite{f4,f5,f6,g2,g3,g10} (see also \cite{AR00}).
 
In the Tully's NBG Catalogue \cite{t3} the inclination angle was calculated
according to the formula:  $i=cos^{-1}{(q^2 -q^2_0 )/(1-q^2_0)}^{-1/2}+3^0$,
where $q=d/D$ is the ratio of the minor to the major axis diameters and
$q_0$  is  "true" axial ratio.  Tully used a standard value $q_0=0.2$.
Worthy of note is that above formula is the modified Holmberg's \cite{Holmberg46}
formula for oblate spheroids. For each galaxy, two angles are
determined: $\delta_D$ - the angle between the normal to the galaxy plane
and the main plane of the coordinate system, and $\eta$ - the angle between
the projection of this normal onto the main plane and the direction towards
to the zero initial meridian. Using the Supergalactic coordinate system
(Flin \& God{\l}owski \cite{f4} based on \cite{TS76}) the following relations
between angles ($L$, $B$, $P$) and ($\delta_D$, $\eta$) are hold:
\begin{equation}
\sin\delta_D  =  -\cos{i}\sin{B} \pm \sin{i}\cos{r}\cos{B},
\end{equation}
\begin{equation}
\sin\eta  =  (\cos\delta_D)^{-1}[-\cos{i}\cos{B}\sin{L} + \sin{i}
(\mp \cos{r}\sin{B}\sin{L} \pm \sin{r}\cos{L})],
\end{equation}
\begin{equation}
\cos\eta  =  (\cos\delta_D)^{-1}[-\cos{i}\cos{B}\cos{L} + \sin{i}
(\mp \cos{r}\sin{B}\cos{L} \mp \sin{r}\sin{L})],
\end{equation}
where $r=P-\pi/2$.
 
In order to detect non-random effects in the distribution of the
investigated angles: $\delta_D$, $\eta$ and $P$ we divided the entire
range of the analyzed angles into $18$ bins and carried
out three different statistical tests. These tests were : the $\chi^2$ test,
the autocorrelation test and the Fourier test  \cite{h4,g2,g3,g10a,g11a}.
 
Let $N$ denotes the total number of galaxies in the considered cluster, and
$N_k$ - the number of galaxies with orientations within the $k$-th  angular
bin. Moreover, $N_{0,k}$ denotes the expected number of galaxies in the $k$-th
bin. In our case all $N_{0,k}$ are equal $N_0$, which is also mean
number of galaxies per bin.
 
Our first test is the $\chi^2$ test:
\begin{equation}
\chi^2 = \sum_{k = 1}^n {(N_k -N\,p_k)^2 \over N\,p_k}= \sum_{k = 1}^n {(N_k -N_{0,k})^2 \over N_{0,k}},
\end{equation}
where $p_k$ is a probability that chosen galaxy falls into $k$th bin.
We divided entire range of a $\theta$ angle into $n$ bins, which gives   $(n-1)$ 
degrees of freedom in the $\chi^2$ test. It means that the expected
value $E(\chi^2)=n-1$ while variance $\sigma^2(\chi^2)=2(n-1)$.
For $n=18$ the $\chi^2$ test yields a critical value $27.59$ (at the significance
level $\alpha=0.05$).

The second auto-correlation test quantifies the correlations between
galaxy numbers in neighboring angle bins. The measure of the correlation
is defined as:
\begin{equation}
C\, = \, \sum_{k = 1}^n { (N_k -N_{0,k})(N_{k+1} -N_{0,k+1} )
\ \over \left[ N_{0,k} N_{0,k+1}\right]^{1/2} },
\end{equation}
where $N_{n+1}=N_1$. Hawley and Peebles \cite{h4} noted that in the case of an
isotropic distribution, we expect $C = 0$ with the standard deviation
$\sigma(C) = n^{1/2}$. God{\l}owski \cite{g11a}, analysing  case that all
$N_{0,k}$ were equal, showed  that in these case the expected value
$E(C)=-1$, while critical value for autocorrelation test was
$C_{cr}\approx 6.89$. The latter value was obtained from numerical simulations
using the method described by God{\l}owski \cite{g11a}.
 
If deviation from isotropy is a slowly varying function of the angle $\theta$
one can use the Fourier test \cite{h4,g2,g3}:
\begin{equation}
N_k = N_{0,k} (1+\Delta_{11} \cos{2 \theta_k} +\Delta_{21} \sin{2\theta_k}),
\end{equation}
we obtain the following expressions for the $\Delta_{i1}$ coefficients:
\begin{equation}
\Delta_{11} = {\sum_{k = 1}^n (N_k -N_{0,k})\cos{2 \theta_k} \over
\sum_{k = 1}^n N_{0,k} \cos^2{2 \theta_k}},
\end{equation}
\begin{equation}
\Delta_{21} = { \sum_{k = 1}^n (N_k-N_{0,k})\sin{2 \theta_k} \over
\sum_{k = 1}^n N_{0,k} \sin^2{2 \theta_k}}.
\end{equation}
Standard deviation of $\sigma(\Delta_{11})$ and $\sigma(\Delta_{12})$ is given
by expressions:
\begin{equation}
\label{eq:f4}
\sigma(\Delta_{11}) =
 \left( {\sum_{k = 1}^n N_{0,k} \cos^2{2 \theta_k} } \right)^{-1/2} \approx
 \left( {2 \over n N_0} \right)^{1/2},
\end{equation}
\begin{equation}
\label{eq:f5}
\sigma(\Delta_{21}) =
 \left( {\sum_{k = 1}^n N_{0,k} \sin^2{2 \theta_k}} \right)^{-1/2} \approx
 \left( {2 \over n N_0} \right)^{1/2}.
\end{equation}
The probability that the amplitude
\begin{equation}
\Delta_1 = \left( \Delta_{11}^2 + \Delta_{21}^2 \right)^{1/2}
\end{equation}
is greater than a certain chosen value is given by the formula:
\begin{equation}
P(>\Delta_1 ) = \exp{\left( -{n \over 4} N_0 \Delta_1^2 \right)}.
\end{equation}
This test was substantially improved by God{\l}owski
 for the case when higher Fourier mode is taken into account:
\begin{equation}
N_k = N_{0,k} (1+\Delta_{11} \cos{2 \theta_k} +\Delta_{21} \sin{2
\theta_k}+\Delta_{12} \cos{4 \theta_k}+\Delta_{22} \sin{4\theta_k}+.....).
\end{equation}
In this case the amplitude $\Delta$ instead $\Delta_1$ is considered.
When we investigate simple case of position angles distribution, $\Delta^2$
is given by simple formula:
$\Delta^2 = \Delta_{11}^2 + \Delta_{21}^2+\Delta_{12}^2 + \Delta_{22}^2$
(see \cite{g3,g11a} for details).
 
The isotropy of the
resultant distributions of the investigated angles was also analyzed by
Kolmogorov-Smirnov test (K-S test). We assumed that the theoretical, random
distribution contains the same number of objects as the observed one. In order
to reject the $H_0$ hypothesis, that the distribution is random one, the value
of observed statistics $\lambda$ should be greater than $\lambda_{cr}= 1.358$
(for $\alpha=0.05$). However one should note, that especially in the case of
position angles, the number of analyzed galaxies is sometimes small and does not
satisfy theoretical tests conditions. That is the reason we repeated our
analysis with different numbers of bins, and revealed insignificant differences
in these cases.

\section{The results}
 
The results were presented in the Tables 1-20. At first,
we analyzed the orientation of galaxies in Tully's groups using, for obtaining
$\delta_D$ and $\eta$ angles, inclinations angles taken directly from NBG
Catalogue \cite{t3} (sample A). These results were shown in the Tables 1-9.
 
For the sample of $All$ galaxies the
analysis of the supergalactic position angles has shown that only one group
($61$) exhibits the alignment of galaxies. Analysis of the distribution of the
angles  giving spatial orientation of galaxies ($\delta_D$ and $\eta$) seems
to show a weak alignment. For $\delta_D$ angle the tests showed that distributions
can be non random in the case of the clusters $11$, $31$, $51$ $41$ and $52$.
For $\eta$ angle we found a possible alignment in the case of clusters $11$, $12$,
$41$, $52$, $64$, $31$ and $51$.
However, in the paper by God{\l}owski \cite{g11b} it was shown that this
alignment is disappeared when we avoid the assumption that the "true" axial ratio
is $q_0=0.2$, which is a rather poor approximation, especially for non-spiral
galaxies. Because of the NBG Catalogue contains morphological types of galaxies
it allowed us to use different values of $q_0$ depending on morphological type
\cite{hhv}. With help by Fouque \& Paturel \cite{fp85} formulae, which
convert $q$ to the standard photometrical axial ratios, the new inclination
angle $i$ for all galaxies in NGB catalogue was computed. The results of our
investigations for that "new" sample of galaxies (sample B) are presented in
Tables 10-18. Please note that above procedure do not change the position
angles of galaxies, instead of the case of position angle $P$, sample B
is related to only those galaxies were the certainly measured angles were
taking into account.
 
Returning to the analysis of samples $A$ divided according to morphological
type, we do not find any alignment for spiral ($SP$) galaxies while for
sample of non-spiral ($NSP$) galaxies we found non randomness in the
distributions of $\delta_D$ and $\eta$ angles for fifth and seventh groups
respectively. During analysis of the sample B we observed disappearing of
the alignment. For position angles $P$ we do not observe any alignment.
During analysis the $\delta_D$ angle for spiral galaxies we observed an 
alignment for the group $53$ while for non-spiral we observed alignment
for groups $11$ and $51$. In the case of the $\eta$ angle the non randomness was
observed in two groups for $All$ galaxies (groups $11$ and $41$) and in four
groups ($11$, $14$, $42$ and $52$) for non-spiral galaxies. It confirms
God{\l}owski' \cite{g11b} conclusion that any possibly observed alignment for
the galaxies in sample $A$ is caused by the reason of the wrong assumption
the "true" axial ratio especially for non-spiral galaxies.

For more detailed analysis we used the method described in \cite{g11a}.
Our question is, whether we could say that we found an alignment
in the whole analyzed sample of $18$ Tully's groups of galaxies or not. So
we computed the mean value and the variance of analyzed statistics:
$\chi^2$, $\Delta_1/\sigma(\Delta_1)$, $\Delta/\sigma(\Delta)$
(i.e. the same statistics were analyzed in \cite{g10a}) for our sample of
$18$ groups and compared them with the results of numerical simulations. We
performed 1000 simulations of 18 fictious clusters, each with a number of
randomly oriented, galaxy's members, the same as in real clusters separately
for $All$ (see \cite{g11b}), for spiral, and for non-spiral galaxies. In the
Table 19 we presented  average values of the analyzed statistics, their standard 
deviations, standard deviations in the sample as well as their standard deviations 
for distribution of $P$ angles, obtained from numerical simulations . One should 
note, that there are some differences in the results of numerical simulations  
for $P$, $\delta_D$ and $\eta$ angles but it does not change our further conclusions.
 
The mean values and variance of analyzed statistics  for sample of real
clusters were presented in the Table 20. It shows that (for sample A)
analysis of the position angles for $All$ and spiral galaxies does not show
a significant deviation from the values expected in the case of random
distributions, while marginal effect is observed for non-spiral galaxies.
The analysis  of $\delta_D$ and $\eta$ angles shows the existence of alignment
at the $2\sigma$ level (with exception of $\Delta/\sigma(\Delta)$ statistics
for $\delta_D$ angle) for subsample of $All$ galaxies and even stronger
effect for non-spiral galaxies. The spiral galaxies do not show this effect.
For the sample $B$ generally we do not observe the deviation from the values
expected in the case of random distributions. The above results allowed us
to conclude that we did not observe any significant alignment for Tully's
groups of galaxies.

\section{Discussion and conclusions}

In the present paper an investigation of the orientation of galaxies inside  $18$
Tully's groups of galaxies belonging to the Local Supercluster divided
according to morphological type was performed. We do not find any significant
alignment in the orientation of galaxies in the analyzed groups of galaxies.
As a result we concluded that the orientations of galaxies in the Tully's groups
are random. Presence of a possible weak alignment for spiral galaxies needs
more investigations in the future.
We also analyzed observational effect generated by the process of
deprojection of  galaxies found by God{\l}owski and Ostrowski \cite{g5},
which masks to the high degree any possible alignment during analysis of the
spatial orientation of galaxies in clusters. We confirmed God{\l}owski's
suggestion \cite{g11,g11b} that this effect is because of wrong aproximation
of the "true shape" of galaxies, especialy for non-spiral galaxies.
We've shown that using "true shape"
of galaxies $q_0$ depending on morphological type according to Heidmann et al.
\cite{hhv} with help of Fouque \& Paturel \cite{fp85} corrections of $q$
to standard photometrical axial ratios, allowed us to avoid this problem. This
gives much more powerful investigation of the spatial orientation of galaxies.
 
Our results, lack of the alignment for less masive galaxies structures together
with results of our previous papers \cite{g4,Baier03,g10,g11,g11a}
confirms our suggestion that alignment of galaxies increases with the mass of 
the structures \cite{g05,g11}. Similar result was also obtain by Aryal et al. \cite{Aryal07} 
based on the series of theirs papers \cite{Aryal04,Aryal05,Aryal06}. 
In our opinion the observed relation between the richness of galaxy cluster and
 the alignment is due to tidal torque, as suggested by \cite{Catelan96}, however 
it is also in agreement with prdiction of the Li model \cite{Li98} in which galaxies 
form in the  rotating universe.

It should be noticed that Gonzalez and Teodoro \cite{Gon10}  interpreted the 
alignment of just the brightest galaxies within a cluster as an effect of action 
of gravitational tidal forces. Recently,  there  have  been  also some  attempts 
to  investigate galaxy  angular  momenta  on a large scale. Paz,  Stasyszyn  and Padilla \cite{Paz08} 
analysing galaxies from the  Sloan  Digital  Sky Survey  catalogue  found  that
the  galaxy  angular  momenta  are aligned  perpendicularly to the planes of 
large-scale structures, while  there is no such effect for the low-mass structures. 
They interpret this as consistent with their simulations based on  the
mechanism  of  tidal interactions. Jones, van  der  Waygaert  and
Aragon-Calvo \cite{Jones10}  found that the  spins  of  spiral  galaxies located 
within cosmic web filaments tend to be aligned along  the larger  axis of the filament, 
which they interpreted as  "fossil" evidence indicating that the action of large 
scale tidal  torques effected the alignments of galaxies located in cosmic filaments.

In the commonly accepted $\Lambda$CDM model, the Universe deems to be spatially
flat, as well as homogeneous and isotropic at appropriate scale. In this model 
the structure were formed from the primordial adiabatic, nearly scale invariant 
Gaussian random fluctuations \cite{Silk68,Peebles70,Sunyaew70}. This picture 
is in agreement with both the numerous numerical simulations \cite{Springel05,w1,w2,Codis12} 
and the observations.

Usually, dependence between the angular momentum and the mass of the
structure is presented as the empirical relation $J\sim M^{5/3}$
\cite{Wesson79,Wesson83,Carrasco82,Brosche86,Paz08,Rom12}, for rewiev see also \cite{Sch09}. 
In our opinion, it is due to tidal torque, as suggested by \cite{HP88,Catelan96}. Also 
\cite{Noh06a,Noh06b} analysing  of the linear tidal torque theory
noticed the connection of the alignment with the considered scale of structure.
However, another possibilities, as Li model \cite{Li98} 
for example, are also not excluded.

\begin{table}
\noindent
\caption{Test for isotropy of the orientations of galaxy plane. The distribution
of the angle $\delta$ of galaxies, inclination was taken directly from NGC Catalogue.}
\begin{tabular}{cccccccc}
angle&group&$\chi^2$&$C$&$P(\Delta_1)$&$P(\Delta)$&$\lambda$&$\Delta_{11}/\sigma(\Delta_{11}$)\\
         & $11$ &  62.8&   9.50& .000& .000&  1.20& -4.09\\
         & $12$ &  25.7& -11.48& .299& .634&  0.55& -1.56\\
         & $13$ &  29.7&  -6.94& .915& .944&  0.62&  0.02\\
         & $14$ &  24.0&   3.52& .170& .254&  0.78& -1.69\\
         & $15$ &  13.1&  -1.88& .791& .729&  0.59& -0.03\\
         & $17$ &  13.8&  -5.50& .621& .743&  0.82& -0.61\\
         & $21$ &  14.2&   0.24& .120& .299&  1.14& -0.38\\
         & $22$ &   7.0&   0.68& .546& .857&  0.62&  0.75\\
 $\delta$& $23$ &  17.0&  -1.82& .613& .701&  0.47&  0.96\\
         & $31$ &  33.7&  15.98& .000& .002&  1.79&  1.36\\
         & $41$ &  22.8&   6.54& .049& .012&  1.37& -0.04\\
         & $42$ &  24.7&  -3.78& .401& .444&  0.75&  0.58\\
         & $44$ &  26.7&   7.95& .047& .050&  1.34& -1.22\\
         & $51$ &  29.6&  -0.41& .025& .018&  1.44&  0.96\\
         & $52$ &  21.6&   3.04& .011& .040&  1.42& -1.81\\
         & $53$ &  13.3&  -3.29& .556& .821&  0.62& -0.60\\
         & $61$ &  19.9&  -3.13& .834& .175&  0.62& -0.54\\
         & $64$ &  28.7&  -7.54& .180& .488&  0.84&  0.55\\
\end{tabular}
\end{table}
 
\begin{table}
\noindent
\caption{Test for isotropy of the orientations of galaxy plane. The distribution
of the angle $\delta$ of galaxies, SP galaxies, inclination was taken directly from NGC Catalogue.}
\begin{tabular}{cccccccc}
angle&group&$\chi^2$&$C$&$P(\Delta_1)$&$P(\Delta)$&$\lambda$&$\Delta_{11}/\sigma(\Delta_{11}$)\\
         & $11$ &  27.6&  -1.85& .278& .517&  0.68&  -1.59\\
         & $12$ &  21.3&  -9.50& .582& .844&  0.41&  -1.01\\
         & $13$ &  30.4&  -9.00& .314& .281&  0.69&   1.34\\
         & $14$ &  15.4&  -1.25& .907& .635&  0.43&   0.37\\
         & $15$ &  12.9&  -1.51& .595& .729&  0.72&   0.79\\
         & $17$ &  13.6&  -8.66& .947& .975&  0.46&  -0.13\\
         & $21$ &  12.3&  -0.52& .636& .403&  0.64&   0.17\\
         & $22$ &  17.7&  -0.69& .063& .197&  1.12&   1.63\\
 $\delta$& $23$ &  15.7&  -1.34& .377& .393&  0.77&   1.18\\
         & $31$ &  22.9&   4.52& .024& .105&  1.27&   1.07\\
         & $41$ &  22.9&   4.91& .487& .033&  0.80&   0.18\\
         & $42$ &  20.2&  -4.17& .807& .303&  0.60&   0.58\\
         & $44$ &  12.9&   0.83& .371& .453&  0.63&  -1.04\\
         & $51$ &  16.6&  -4.99& .729& .582&  0.55&   0.53\\
         & $52$ &  19.0&  -1.47& .057& .213&  1.07&  -1.59\\
         & $53$ &  19.7&  -1.92& .120& .196&  1.05&  -1.20\\
         & $61$ &  19.0&  -5.70& .565& .392&  0.65&   0.37\\
         & $64$ &  20.7&  -2.31& .789& .668&  0.37&   0.29\\
\end{tabular}
\end{table}

\begin{table}
\noindent
\caption{Test for isotropy of the orientations of galaxy plane. The distribution
of the angle $\delta$ of galaxies, NSP galaxies, inclination was taken directly from NGC Catalogue.}
\begin{tabular}{cccccccc}
angle&group&$\chi^2$&$C$&$P(\Delta_1)$&$P(\Delta)$&$\lambda$&$\Delta_{11}/\sigma(\Delta_{11}$)\\
         & $11$ &  93.3&   4.00& .000& .000&  1.80&  -5.10\\
         & $12$ &  26.6&  -5.46& .362& .481&  0.75&  -1.35\\
         & $13$ &  34.6&  -1.84& .274& .125&  0.69&  -1.60\\
         & $14$ &  26.3&   8.14& .011& .689&  1.07&  -2.67\\
         & $15$ &  21.5&   4.03& .027& .060&  1.48&  -1.34\\
         & $17$ &  11.9&   0.45& .526& .688&  0.72&  -0.79\\
         & $21$ &  32.9&  -3.95& .038& .077&  1.58&  -1.01\\
         & $22$ &  15.3&  -1.46& .440& .396&  0.96&  -0.91\\
 $\delta$& $23$ &  13.3&  -1.20& .735& .724&  0.73&  -0.10\\
         & $31$ &  27.0&   5.30& .007& .468&  1.45&   0.87\\
         & $41$ &  18.1&   1.02& .038& .491&  1.39&  -0.31\\
         & $42$ &  21.8&   2.01& .195& .472&  1.07&   0.17\\
         & $44$ &  22.5&   7.63& .052& .126&  1.40&  -0.64\\
         & $51$ &  33.3&   7.58& .001& .033&  2.05&   0.94\\
         & $52$ &  22.9&   2.63& .160& .058&  1.23&  -0.90\\
         & $53$ &  32.1&  14.22& .000& .136&  2.01&   0.73\\
         & $61$ &  23.6&   0.99& .094& .035&  1.03&  -1.77\\
         & $64$ &  23.2&  -0.18& .090& .453&  1.17&   0.53\\
\end{tabular}
\end{table}

\begin{table}
\noindent
\caption{Test for isotropy of the orientations of galaxy plane. The distribution
of the angle $\eta$ of galaxies, inclination was taken directly from NGC Catalogue.}
\begin{tabular}{cccccccc}
angle&group&$\chi^2$&$C$&$P(\Delta_1)$&$P(\Delta)$&$\lambda$&$\Delta_{11}/\sigma(\Delta_{11}$)\\
         & $11$ &  60.0&   5.96& .000&  .000&  2.00& -4.09\\
         & $12$ &  28.5&   7.56& .001&  .010&  1.70& -1.56\\
         & $13$ &  25.6&   2.78& .079&  .105&  0.77&  0.02\\
         & $14$ &  27.4&   6.63& .090&  .148&  1.00& -1.69\\
         & $15$ &  22.9&   2.51& .764&  .209&  0.91& -0.03\\
         & $17$ &  13.1&  -5.97& .470&  .820&  0.46& -0.61\\
         & $21$ &  26.9&  -3.55& .054&  .202&  1.23& -0.38\\
         & $22$ &  11.7&   5.57& .177&  .224&  0.98&  0.75\\
 $\eta$  & $23$ &  20.2&  -0.10& .081&  .232&  1.09&  0.96\\
         & $31$ &  24.0&   0.43& .046&  .036&  1.56&  1.36\\
         & $41$ &  27.2&  10.22& .001&  .002&  1.95& -0.04\\
         & $42$ &  15.9&   3.30& .033&  .069&  1.24&  0.58\\
         & $44$ &  20.4&  -3.05& .226&  .138&  0.78& -1.22\\
         & $51$ &  30.6&  -5.37& .042&  .125&  1.15&  0.96\\
         & $52$ &  38.1&  12.29& .001&  .001&  1.92& -1.81\\
         & $53$ &  12.2&  -8.28& .816&  .949&  0.34& -0.60\\
         & $61$ &  23.7& -12.56& .549&  .511&  0.64& -0.54\\
         & $64$ &  50.1&  -3.53& .002&  .000&  1.88&  0.55\\
\end{tabular}
\end{table}
 
\begin{table}
\noindent
\caption{Test for isotropy of the orientations of galaxy plane. The distribution
of the angle $\eta$ of galaxies, SP galaxies, inclination was taken directly from NGC Catalogue.}
\begin{tabular}{cccccccc}
angle&group&$\chi^2$&$C$&$P(\Delta_1)$&$P(\Delta)$&$\lambda$&$\Delta_{11}/\sigma(\Delta_{11}$)\\
         & $11$ &  32.1&  -3.44& .070&  .043&  1.17&   2.30\\
         & $12$ &  18.7&   2.93& .099&  .133&  1.17&  -0.17\\
         & $13$ &  25.8&   1.92& .519&  .115&  1.16&   1.04\\
         & $14$ &  12.9&   0.37& .372&  .356&  0.71&  -1.38\\
         & $15$ &  25.5&   2.89& .537&  .168&  0.97&   0.63\\
         & $17$ &   8.0&  -2.57& .707&  .948&  0.67&   0.48\\
         & $21$ &  14.6&   1.00& .216&  .424&  0.89&   1.12\\
         & $22$ &  16.3&   3.92& .481&  .142&  0.87&   1.10\\
 $\eta$  & $23$ &  14.5&  -2.73& .777&  .884&  0.44&  -0.34\\
         & $31$ &  22.7&   1.27& .181&  .079&  1.38&   1.19\\
         & $41$ &  16.3&   6.14& .023&  .069&  1.34&   2.48\\
         & $42$ &  13.4&  -6.50& .712&  .947&  0.48&  -0.09\\
         & $44$ &  16.5&  -5.62& .308&  .427&  0.59&   1.17\\
         & $51$ &  34.8&  -6.10& .421&  .603&  0.98&   0.69\\
         & $52$ &  25.8&   4.05& .043&  .092&  1.19&   1.06\\
         & $53$ &  16.6&  -8.13& .423&  .623&  0.53&   0.30\\
         & $61$ &  24.5&  -7.82& .343&  .159&  1.14&   0.83\\
         & $64$ &  20.2&  -5.73& .339&  .281&  0.86&   1.21\\
\end{tabular}
\end{table}
 
\begin{table}
\noindent
\caption{Test for isotropy of the orientations of galaxy plane. The distribution
of the angle $\eta$ of galaxies, NSP galaxies, inclination was taken directly from NGC Catalogue.}
\begin{tabular}{cccccccc}
angle&group&$\chi^2$&$C$&$P(\Delta_1)$&$P(\Delta)$&$\lambda$&$\Delta_{11}/\sigma(\Delta_{11}$)\\
         & $11$ &  67.2&  32.17& .000&  .000&  1.92&   6.53\\
         & $12$ &  26.6&  12.58& .002&  .003&  1.40&  -1.45\\
         & $13$ &  22.1&  -1.12& .097&  .164&  1.19&   1.78\\
         & $14$ &  52.4&  16.75& .001&  .001&  1.53&   2.45\\
         & $15$ &  13.0&   3.00& .879&  .264&  0.50&  -0.46\\
         & $17$ &  18.9&  -0.12& .019&  .091&  1.24&  -1.14\\
         & $21$ &  32.6& -10.03& .066&  .195&  0.89&  -0.58\\
         & $22$ &   6.4&   1.35& .309&  .523&  0.56&   1.09\\
 $\eta$  & $23$ &  22.5&   3.08& .006&  .036&  1.39&  -1.71\\
         & $31$ &  13.8&   0.27& .139&  .407&  0.91&   1.91\\
         & $41$ &  26.7&   2.18& .019&  .047&  1.48&   1.59\\
         & $42$ &  30.5&  13.91& .001&  .000&  1.89&   0.87\\
         & $44$ &  23.4&  -3.57& .551&  .390&  0.94&  -0.01\\
         & $51$ &  25.2&   6.91& .006&  .013&  0.96&   3.21\\
         & $52$ &  27.2&  14.91& .009&  .001&  1.78&   2.08\\
         & $53$ &   8.3&  -1.85& .695&  .899&  0.38&  -0.31\\
         & $61$ &  22.8&   0.00& .012&  .048&  1.38&   0.63\\
         & $64$ &  47.0&   5.50& .001&  .000&  2.00&   3.19\\
\end{tabular}
\end{table}
 
\begin{table}
\noindent
\caption{Test for isotropy of the orientations of galaxy plane. The distribution
of the angle $P$ of galaxies.}
\begin{tabular}{cccccccc}
angle&group&$\chi^2$&$C$&$P(\Delta_1)$&$P(\Delta)$&$\lambda$&$\Delta_{11}/\sigma(\Delta_{11}$)\\
         & $11$ &  22.7& -11.42& .728&  .756&  0.53&   0.78\\
         & $12$ &  17.6&  -1.57& .198&  .302&  0.88&  -0.61\\
         & $13$ &  13.4&  -2.48& .714&  .621&  0.41&  -0.28\\
         & $14$ &  14.9&  -1.05& .990&  .204&  0.49&  -0.09\\
         & $15$ &  11.3&  -0.75& .185&  .382&  1.06&  -0.03\\
         & $17$ &  10.7&  -5.64& .727&  .734&  0.62&  -0.50\\
         & $21$ &  13.5&  -2.84& .878&  .936&  0.67&  -0.38\\
         & $22$ &  16.0&  -1.98& .910&  .895&  0.63&  -0.41\\
 $P$     & $23$ &  12.3&   0.27& .230&  .453&  0.81&  -1.37\\
         & $31$ &  20.1&   1.00& .729&  .204&  0.82&   0.46\\
         & $41$ &  20.0&  10.67& .595&  .005&  1.09&  -0.58\\
         & $42$ &  19.0&   1.51& .124&  .222&  0.91&  -1.51\\
         & $44$ &  18.9&   1.64& .367&  .726&  0.98&   0.57\\
         & $51$ &  23.1&   3.00& .576&  .045&  0.88&  -1.04\\
         & $52$ &  14.8&  -0.32& .631&  .798&  0.77&   0.77\\
         & $53$ &  17.5&   2.82& .080&  .192&  1.36&   1.61\\
         & $61$ &  27.9&   6.48& .007&  .002&  1.68&  -0.76\\
         & $64$ &  18.4&   2.10& .295&  .419&  0.91&  -1.56\\
\end{tabular}
\end{table}
 
\begin{table}
\noindent
\caption{Test for isotropy of the orientations of galaxy plane. The distribution
of the angle $P$ of galaxies, SP galaxies.}
\begin{tabular}{cccccccc}
angle&group&$\chi^2$&$C$&$P(\Delta_1)$&$P(\Delta)$&$\lambda$&$\Delta_{11}/\sigma(\Delta_{11}$)\\
         & $11$ &  21.8&  -9.88& .821&  .853&  0.50&  -0.18\\
         & $12$ &  14.1&   0.55& .173&  .339&  0.86&  -0.54\\
         & $13$ &  15.6&   4.80& .430&  .073&  0.79&  -1.21\\
         & $14$ &  16.3&   4.30& .326&  .167&  0.74&  -0.65\\
         & $15$ &  12.2&  -2.00& .294&  .566&  0.96&  -0.38\\
         & $17$ &  11.7&  -5.00& .702&  .733&  0.62&   0.00\\
         & $21$ &  22.2&  -8.00& .577&  .636&  0.93&  -1.03\\
         & $22$ &  16.3&  -1.67& .416&  .717&  0.77&  -1.28\\
 $P$     & $23$ &  12.8&   1.69& .272&  .270&  0.78&  -0.74\\
         & $31$ &  14.1&  -2.12& .532&  .416&  0.57&  -0.26\\
         & $41$ &  21.7&  10.32& .972&  .003&  0.78&  -0.07\\
         & $42$ &  14.7&   1.86& .146&  .194&  0.80&  -1.02\\
         & $44$ &  16.0&  -3.00& .371&  .700&  0.71&   0.46\\
         & $51$ &  23.9&  -0.67& .686&  .041&  0.84&  -0.84\\
         & $52$ &  12.2&  -2.95& .533&  .851&  0.56&   0.99\\
         & $53$ &  16.4&  -3.55& .508&  .613&  0.78&   0.66\\
         & $61$ &  21.0&   1.51& .060&  .036&  1.25&  -0.43\\
         & $64$ &  16.0&   0.70& .683&  .344&  1.07&  -0.66\\
\end{tabular}
\end{table}
 
\begin{table}
\noindent
\caption{Test for isotropy of the orientations of galaxy plane. The distribution
of the angle $P$ of galaxies, NSP galaxies.}
\begin{tabular}{cccccccc}
angle&group&$\chi^2$&$C$&$P(\Delta_1)$&$P(\Delta)$&$\lambda$&$\Delta_{11}/\sigma(\Delta_{11}$)\\
         & $11$ &  22.7&   4.66& .044&  .031&  0.81&   1.99\\
         & $12$ &  18.3&  -1.08& .926&  .931&  0.41&  -0.27\\
         & $13$ &  14.2&  -2.90& .475&  .432&  0.57&   1.04\\
         & $14$ &  19.3&  -4.94& .399&  .270&  0.57&   0.54\\
         & $15$ &  18.8&  -2.80& .451&  .326&  1.02&   0.68\\
         & $17$ &  14.5&  -3.50& .703&  .769&  0.63&  -0.83\\
         & $21$ &  16.9&  -7.35& .604&  .846&  0.36&   0.96\\
         & $22$ &  20.0&  -1.38& .600&  .798&  0.61&   0.99\\
 $P$     & $23$ &  26.4&  -7.00& .269&  .379&  0.65&  -1.54\\
         & $31$ &  27.2&   3.62& .250&  .148&  1.34&   1.51\\
         & $41$ &  11.0&  -0.25& .312&  .526&  0.86&  -0.96\\
         & $42$ &  19.8&  -1.80& .396&  .743&  0.65&  -1.34\\
         & $44$ &  16.1&   0.71& .898&  .844&  0.84&   0.34\\
         & $51$ &  14.3&   1.43& .585&  .847&  0.62&  -0.63\\
         & $52$ &  30.0&   6.00& .104&  .045&  1.73&  -0.19\\
         & $53$ &  17.3&   4.18& .039&  .164&  1.37&   2.08\\
         & $61$ &  16.9&   5.29& .051&  .067&  1.34&  -0.90\\
         & $64$ &  20.1&   0.45& .168&  .321&  0.80&  -1.74\\
\end{tabular}
\end{table}

\begin{table}
\noindent
\caption{Test for isotropy of the orientations of galaxy plane. The distribution
of the angle $\delta$ of galaxies, inclination was obtained according to HHV and FP corrections.}
\begin{tabular}{cccccccc}
angle&group&$\chi^2$&$C$&$P(\Delta_1)$&$P(\Delta)$&$\lambda$&$\Delta_{11}/\sigma(\Delta_{11}$)\\
         & $11$ &  14.3&  -2.46& .504&  .848&  0.39&  -1.17\\
         & $12$ &  12.4&   2.49& .749&  .680&  0.37&   0.74\\
         & $13$ &   8.7&   0.86& .403&  .694&  0.54&   1.25\\
         & $14$ &  22.7&   1.50& .658&  .466&  0.52&  -0.89\\
         & $15$ &  26.2&   4.40& .038&  .034&  1.25&   1.90\\
         & $17$ &   7.0&  -1.57& .959&  .917&  0.48&  -0.11\\
         & $21$ &  19.7&   6.05& .615&  .049&  0.85&   0.72\\
         & $22$ &  17.9&   1.77& .062&  .128&  1.30&   1.44\\
 $\delta$& $23$ &  22.1&  -3.00& .155&  .266&  0.71&   1.88\\
         & $31$ &  19.1&   3.90& .095&  .245&  0.74&   0.50\\
         & $41$ &  20.9&  -4.77& .374&  .476&  0.69&   1.26\\
         & $42$ &  18.3&  -4.08& .462&  .293&  0.86&  -0.95\\
         & $44$ &  23.9&   0.74& .581&  .115&  0.93&  -0.63\\
         & $51$ &  18.3&   3.40& .826&  .851&  0.61&   0.32\\
         & $52$ &  20.7&   4.97& .013&  .043&  1.53&  -1.51\\
         & $53$ &  13.2&  -1.46& .696&  .720&  0.62&  -0.67\\
         & $61$ &  25.3&  -1.26& .452&  .376&  1.03&  -0.22\\
         & $64$ &  25.9&  -1.99& .227&  .354&  0.78&   0.06\\
\end{tabular}
\end{table}
 
\begin{table}
\noindent
\caption{Test for isotropy of the orientations of galaxy plane. The distribution
of the angle $\delta$ of galaxies, SP galaxies, inclination was obtained according to HHV and FP corrections.}
\begin{tabular}{cccccccc}
angle&group&$\chi^2$&$C$&$P(\Delta_1)$&$P(\Delta)$&$\lambda$&$\Delta_{11}/\sigma(\Delta_{11}$)\\
         & $11$ &  19.7&  -4.98& .894& .402&  0.44&   0.36\\
         & $12$ &  19.0&  -5.69& .979& .716&  0.51&   0.20\\
         & $13$ &  29.4&   1.55& .085& .066&  0.85&   2.01\\
         & $14$ &  17.8&   1.49& .796& .139&  0.60&   0.65\\
         & $15$ &  18.6&  -4.82& .330& .624&  0.93&   0.96\\
         & $17$ &  10.0&  -3.30& .910& .947&  0.52&   0.20\\
         & $21$ &  19.6&  -3.88& .794& .176&  0.89&   0.41\\
         & $22$ &  14.7&   3.11& .340& .345&  0.89&   1.07\\
 $\delta$& $23$ &  16.0&  -3.46& .258& .359&  0.77&   1.36\\
         & $31$ &  25.1&  -0.24& .170& .284&  0.88&   1.30\\
         & $41$ &  16.7&  -2.21& .616& .229&  0.66&   0.44\\
         & $42$ &  18.5&   1.64& .729& .069&  0.76&   0.80\\
         & $44$ &   7.9&  -1.11& .583& .869&  0.46&  -0.94\\
         & $51$ &  17.8&  -5.34& .976& .854&  0.29&   0.19\\
         & $52$ &  17.6&   5.95& .010& .054&  1.23&  -1.88\\
         & $53$ &  19.5&   3.85& .014& .036&  1.42&  -1.28\\
         & $61$ &  22.8&  -2.37& .128& .223&  1.06&   1.40\\
         & $64$ &  17.7&   3.38& .930& .442&  0.28&   0.38\\
\end{tabular}
\end{table}

\begin{table}
\noindent
\caption{Test for isotropy of the orientations of galaxy plane. The distribution
of the angle $\delta$ of galaxies, NSP galaxies, inclination was obtained according to HHV and FP corrections.}
\begin{tabular}{cccccccc}
angle&group&$\chi^2$&$C$&$P(\Delta_1)$&$P(\Delta)$&$\lambda$&$\Delta_{11}/\sigma(\Delta_{11}$)\\
         & $11$ & 27.2&  8.78& .006& .025&  1.04& -3.07\\
         & $12$ & 21.0&  6.47& .091& .035&  0.92& -2.18\\
         & $13$ & 11.2&  0.03& .947& .420&  0.51&  0.32\\
         & $14$ & 14.4&  7.68& .613& .115&  0.55& -0.61\\
         & $15$ & 11.9&  0.74& .808& .777&  0.48& -0.03\\
         & $17$ &  6.7&  1.16& .900& .815&  0.39& -0.01\\
         & $21$ & 12.3& -0.57& .802& .947&  0.50& -0.13\\
         & $22$ & 21.0& -1.43& .900& .965&  0.52& -0.30\\
 $\delta$& $23$ & 15.7&  3.03& .698& .662&  0.44&  0.85\\
         & $31$ & 16.4& -3.81& .428& .522&  0.71&  0.23\\
         & $41$ & 17.7& -0.62& .327& .653&  0.90&  1.18\\
         & $42$ & 10.2&  2.47& .384& .745&  0.58&  1.07\\
         & $44$ & 10.2&  1.38& .098& .319&  1.05& -0.49\\
         & $51$ & 11.0&  3.12& .271& .511&  0.84&  0.96\\
         & $52$ &  9.9&  2.15& .435& .727&  0.78&  0.23\\
         & $53$ & 26.5&  9.98& .005& .008&  1.56&  0.67\\
         & $61$ & 17.4&  1.63& .647& .313&  0.62& -0.93\\
         & $64$ & 19.7& -1.93& .226& .443&  0.76&  0.29\\
\end{tabular}
\end{table}
 
\begin{table}
\noindent
\caption{Test for isotropy of the orientations of galaxy plane. The distribution
of the angle $\eta$ of galaxies, inclination was obtained according to HHV and FP corrections.}
\begin{tabular}{cccccccc}
angle&group&$\chi^2$&$C$&$P(\Delta_1)$&$P(\Delta)$&$\lambda$&$\Delta_{11}/\sigma(\Delta_{11}$)\\
         & $11$ &  44.1&  13.00& .000&  .000&  1.37&   3.77\\
         & $12$ &  21.3&   3.87& .075&  .126&  1.65&  -0.90\\
         & $13$ &  18.3&   5.17& .272&  .090&  0.67&   1.53\\
         & $14$ &   8.5&   0.34& .418&  .638&  0.45&   0.55\\
         & $15$ &  22.0& -11.48& .651&  .777&  0.57&   0.72\\
         & $17$ &   8.2&  -3.05& .603&  .852&  0.40&  -0.65\\
         & $21$ &  15.8&  -6.45& .718&  .912&  0.75&   0.17\\
         & $22$ &  14.6&   1.00& .767&  .964&  0.53&   0.73\\
 $\eta$  & $23$ &   7.3&  -1.18& .809&  .956&  0.40&   0.06\\
         & $31$ &  20.2&   0.17& .047&  .103&  0.81&   2.31\\
         & $41$ &  38.3&  20.16& .007&  .000&  1.61&   0.97\\
         & $42$ &  20.6&   4.94& .729&  .023&  0.59&   0.80\\
         & $44$ &  19.5&  -3.95& .763&  .441&  0.55&  -0.08\\
         & $51$ &  20.8&   2.61& .736&  .137&  0.66&   0.71\\
         & $52$ &  17.6&  -0.37& .085&  .276&  0.99&   0.93\\
         & $53$ &  19.3&  -3.50& .888&  .361&  0.51&  -0.06\\
         & $61$ &  23.7&   2.09& .045&  .043&  1.22&   1.39\\
         & $64$ &  23.3&  -5.65& .289&  .291&  0.76&   1.46\\
\end{tabular}
\end{table}
 
\begin{table}
\noindent
\caption{Test for isotropy of the orientations of galaxy plane. The distribution
of the angle $\eta$ of galaxies, SP galaxies, inclination was obtained according to HHV and FP corrections.}
\begin{tabular}{cccccccc}
angle&group&$\chi^2$&$C$&$P(\Delta_1)$&$P(\Delta)$&$\lambda$&$\Delta_{11}/\sigma(\Delta_{11}$)\\
         & $11$ &  25.3&  6.11& .222& .010&  0.77&  1.68\\
         & $12$ &  19.6&  3.51& .673& .264&  0.77& -0.12\\
         & $13$ &  23.5&  0.03& .647& .069&  1.05&  0.91\\
         & $14$ &  16.9&  0.28& .289& .348&  0.78& -1.49\\
         & $15$ &  22.0&  3.85& .331& .097&  1.08&  0.27\\
         & $17$ &  11.1& -1.39& .659& .906&  0.67&  0.37\\
         & $21$ &  14.0& -5.30& .801& .953&  0.45&  0.29\\
         & $22$ &  21.7& -5.08& .914& .230&  0.43&  0.33\\
 $\eta$  & $23$ &  12.1& -1.76& .895& .977&  0.32& -0.47\\
         & $31$ &  24.1&  0.81& .121& .038&  1.48&  0.89\\
         & $41$ &  21.4& -3.86& .149& .156&  0.80&  1.95\\
         & $42$ &   9.0& -3.32& .872& .935&  0.45& -0.27\\
         & $44$ &  13.1& -1.12& .322& .466&  0.62&  1.40\\
         & $51$ &  25.6& -1.04& .867& .368&  0.90&  0.41\\
         & $52$ &  14.0&  1.80& .112& .223&  0.95&  0.07\\
         & $53$ &  14.4& -2.30& .384& .643&  0.61&  0.11\\
         & $61$ &  14.2&  5.27& .067& .068&  1.28&  1.30\\
         & $64$ &  15.3&  0.27& .300& .465&  0.94&  1.33\\
\end{tabular}
\end{table}
 
\begin{table}
\noindent
\caption{Test for isotropy of the orientations of galaxy plane. The distribution
of the angle $\eta$ of galaxies, NSP galaxies, inclination was obtained according to HHV and FP corrections.}
\begin{tabular}{cccccccc}
angle&group&$\chi^2$&$C$&$P(\Delta_1)$&$P(\Delta)$&$\lambda$&$\Delta_{11}/\sigma(\Delta_{11}$)\\
         & $11$ &  29.3&  14.70& .000& .000&  1.27&  4.09\\
         & $12$ &  10.3&  -1.23& .509& .794&  0.40& -0.96\\
         & $13$ &  14.5&  -4.23& .617& .635&  0.60&  0.80\\
         & $14$ &  32.0&  10.08& .004& .002&  1.40&  2.09\\
         & $15$ &  16.0&  -3.50& .808& .915&  0.67& -0.18\\
         & $17$ &  17.9&   7.82& .091& .123&  1.24& -0.64\\
         & $21$ &  14.6&  -4.47& .808& .632&  0.59& -0.53\\
         & $22$ &  12.7&  -2.57& .986& .988&  0.46& -0.15\\
 $\eta$  & $23$ &  12.8&  -0.38& .301& .630&  0.50& -0.89\\
         & $31$ &  14.5&   2.69& .167& .237&  1.20&  0.99\\
         & $41$ &  21.3&   1.91& .052& .124&  1.11&  1.79\\
         & $42$ &  36.5&  16.64& .009& .000&  1.89&  0.26\\
         & $44$ &  15.7&   4.14& .479& .101&  0.82& -1.21\\
         & $51$ &  19.9&   1.88& .098& .197&  1.05&  2.06\\
         & $52$ &  21.5&   6.32& .030& .050&  1.42&  2.05\\
         & $53$ &  12.5&  -3.00& .921& .953&  0.49&  0.03\\
         & $61$ &  14.4&  -1.80& .139& .412&  0.77&  0.37\\
         & $64$ &  27.0&  -6.00& .458& .692&  0.67&  1.24\\
\end{tabular}
\end{table}

\begin{table}
\noindent
\caption{Test for isotropy of the orientations of galaxy plane. The distribution
of the angle $P$ of galaxies. Only galaxies with certain measure $P$ are taken into account.}
\begin{tabular}{cccccccc}
angle&group&$\chi^2$&$C$&$P(\Delta_1)$&$P(\Delta)$&$\lambda$&$\Delta_{11}/\sigma(\Delta_{11}$)\\
         & $11$ &  17.7&  -9.32& .991&  .910&  0.40&  -0.01\\
         & $12$ &  19.5&   0.19& .113&  .137&  0.99&  -0.49\\
         & $13$ &  16.7&   1.40& .805&  .245&  0.81&  -0.63\\
         & $14$ &  14.2&   2.21& .943&  .196&  0.44&  -0.34\\
         & $15$ &  12.0&  -0.86& .201&  .381&  1.03&   0.26\\
         & $17$ &  10.7&  -5.57& .656&  .715&  0.58&  -0.78\\
         & $21$ &  11.2&  -3.53& .859&  .939&  0.55&  -0.36\\
         & $22$ &  15.1&  -1.49& .997&  .999&  0.57&  -0.01\\
 $P$     & $23$ &  12.3&   0.27& .230&  .453&  0.81&  -1.37\\
         & $31$ &  23.3&   4.07& .444&  .052&  0.83&   0.43\\
         & $41$ &  21.3&   9.17& .481&  .011&  1.08&  -0.58\\
         & $42$ &  18.1&   2.38& .075&  .172&  1.04&  -1.59\\
         & $44$ &  19.3&  -2.14& .352&  .694&  0.84&   0.79\\
         & $51$ &  21.0&   0.00& .759&  .069&  0.82&  -0.60\\
         & $52$ &   8.5&  -1.63& .788&  .878&  0.37&   0.68\\
         & $53$ &  16.3&   0.86& .135&  .281&  1.24&   1.32\\
         & $61$ &  27.7&   1.38& .040&  .011&  1.47&  -0.54\\
         & $64$ &  17.5&   3.08& .331&  .519&  1.14&  -1.28\\
\end{tabular}
\end{table}
 
\begin{table}
\noindent
\caption{Test for isotropy of the orientations of galaxy plane. The distribution
of the angle $P$ of galaxies, SP galaxies. Only galaxies with certain measure $P$ are taken into account.}
\begin{tabular}{cccccccc}
angle&group&$\chi^2$&$C$&$P(\Delta_1)$&$P(\Delta)$&$\lambda$&$\Delta_{11}/\sigma(\Delta_{11}$)\\
         & $11$ &  20.1&  -8.40& .570&  .772&  0.52&  -0.58\\
         & $12$ &  16.6&  -0.92& .166&  .345&  0.87&  -0.54\\
         & $13$ &  15.6&   4.80& .430&  .073&  0.79&  -1.21\\
         & $14$ &  15.6&   4.53& .406&  .199&  0.67&  -0.54\\
         & $15$ &  14.0&  -2.50& .275&  .482&  1.00&  -0.39\\
         & $17$ &  11.7&  -5.00& .702&  .733&  0.62&   0.00\\
         & $21$ &  21.4&  -8.66& .657&  .662&  0.88&  -0.82\\
         & $22$ &  13.8&  -1.09& .847&  .987&  0.67&  -0.36\\
 $P$     & $23$ &  12.8&   1.69& .272&  .270&  0.78&  -0.74\\
         & $31$ &  18.2&  -1.39& .266&  .169&  0.74&  -0.45\\
         & $41$ &  20.0&   8.35& .850&  .012&  0.74&  -0.10\\
         & $42$ &  14.6&   1.34& .149&  .236&  0.89&  -0.91\\
         & $44$ &  16.0&  -3.00& .371&  .700&  0.71&   0.46\\
         & $51$ &  21.0&  -1.50& .817&  .075&  0.82&  -0.48\\
         & $52$ &   7.6&   0.17& .376&  .673&  0.55&   1.32\\
         & $53$ &  17.0&  -4.94& .511&  .719&  0.76&   0.61\\
         & $61$ &  22.4&  -2.18& .133&  .078&  1.11&  -0.23\\
         & $64$ &  16.0&   3.00& .506&  .369&  1.18&  -0.69\\
\end{tabular}
\end{table}
 
\begin{table}
\noindent
\caption{Test for isotropy of the orientations of galaxy plane. The distribution
of the angle $P$ of galaxies, NSP galaxies. Only galaxies with certain measure $P$ are taken into account.}
\begin{tabular}{cccccccc}
angle&group&$\chi^2$&$C$&$P(\Delta_1)$&$P(\Delta)$&$\lambda$&$\Delta_{11}/\sigma(\Delta_{11}$)\\
         & $11$ &  15.3&   5.96& .101&  .158&  0.85&   1.32\\
         & $12$ &  14.0&   1.00& .664&  .408&  0.47&   0.00\\
         & $13$ &  11.6&  -1.00& .640&  .889&  0.74&   0.83\\
         & $14$ &  17.4&  -2.17& .350&  .369&  0.82&   0.12\\
         & $15$ &  12.0&   3.00& .180&  .167&  0.95&   1.64\\
         & $17$ &  11.0&  -1.86& .374&  .665&  0.59&  -1.35\\
         & $21$ &  16.0&  -7.00& .759&  .734&  0.71&   0.73\\
         & $22$ &  19.4&  -1.14& .818&  .951&  0.59&   0.44\\
 $P$     & $23$ &  26.4&  -7.00& .269&  .379&  0.65&  -1.54\\
         & $31$ &  30.0&   6.00& .171&  .045&  1.44&   1.83\\
         & $41$ &   9.0&   1.50& .330&  .436&  0.96&  -0.97\\
         & $42$ &  16.8&   2.09& .178&  .331&  0.80&  -1.85\\
         & $44$ &  15.0&   3.00& .613&  .262&  0.96&   0.99\\
         & $51$ &   9.0&   0.00& .916&  .924&  0.38&  -0.40\\
         & $52$ &  27.0&  -3.00& .097&  .155&  0.87&  -1.22\\
         & $53$ &  14.2&   1.60& .115&  .319&  1.17&   1.60\\
         & $61$ &  18.0&   4.50& .180&  .167&  1.15&  -0.83\\
         & $64$ &  16.1&  -1.86& .422&  .648&  0.50&  -1.31\\
\end{tabular}
\end{table}


\begin{table}
\noindent
\caption {The results of numerical simulations for positions angles $P$}.
\begin{tabular}{cccccc}
\hline
Sample&Test&$\bar{x}$&$\sigma(x)$&$\sigma(\bar{x})$&$\sigma(\sigma(x))$\\
\hline
    & $\chi^2$                       & 16.9524&  1.4592&   0.0461&  0.0326 \\
All & $\Delta_{1}/\sigma(\Delta_{1})$&  1.2513&  0.1543&   0.0048&  0.0034 \\
    & $\Delta/\sigma(\Delta)$        &  1.8772&  0.1581&   0.0050&  0.0035 \\
\hline
    & $\chi^2$                       &  17.0283& 1.3628&   0.0431&  0.0305\\
SP  & $\Delta_{1}/\sigma(\Delta_{1})$&   1.2552& 0.1544&   0.0049&  0.0035\\
    & $\Delta/\sigma(\Delta)$        &   1.8827& 0.1543&   0.0049&  0.0035\\
\hline
    & $\chi^2$                       & 17.1042&   1.3485&  0.0426&  0.0302\\
NSP & $\Delta_{1}/\sigma(\Delta_{1})$&  1.2607&   0.1541&  0.0049&  0.0034\\
    & $\Delta/\sigma(\Delta)$        &  1.8900&   0.1595&  0.0050&  0.0036\\
\hline
\end{tabular}
\end{table}
 
\begin{table}
\noindent
\caption {The statistics of the observed distributions for real clusters}
\begin{tabular}{cc|cc|cc|cc}
\hline
\multicolumn{2}{c}{}&
\multicolumn{2}{c}{$P$}&
\multicolumn{2}{c}{$\delta_D$}&
\multicolumn{2}{c}{$\eta$}\\
\hline
Sample&Test&$\bar{x}$&$\sigma(x)$&$\bar{x}$&$\sigma(x)$&$\bar{x}$&$\sigma(x)$\\
\hline
         &$\chi^2$                       & 17.34& 1.06& 23.79& 2.85& 26.58& 2.95\\
$A_{all}$&$\Delta_{1}/\sigma(\Delta_{1})$&  1.28& 0.18&  1.85& 0.26&  2.44& 0.29\\
         &$\Delta/\sigma(\Delta)$        &  2.11& 0.21&  2.33& 0.26&  2.91& 0.27\\
\hline
         &$\chi^2$                       & 16.61& 0.91& 18.93& 1.18& 19.95& 1.65\\
$A_{sp}$ &$\Delta_{1}/\sigma(\Delta_{1})$&  1.26& 0.12&  1.31& 0.16&  1.53& 0.16\\
         &$\Delta/\sigma(\Delta)$        &  2.13& 0.18&  1.98& 0.14&  2.18& 0.17\\
\hline
         &$\chi^2$                       & 19.10& 1.16& 27.79& 4.16& 27.03& 3.61\\
$A_{nsp}$&$\Delta_{1}/\sigma(\Delta_{1})$&  1.45& 0.15&  2.45& 0.27&  2.78& 0.33\\
         &$\Delta/\sigma(\Delta)$        &  1.97& 0.17&  2.87& 0.28&  3.27& 0.33\\
\hline
         &$\chi^2$                       & 16.80& 1.15& 18.70& 1.33& 20.19& 2.17\\
$B_{all}$&$\Delta_{1}/\sigma(\Delta_{1})$&  1.22& 0.18&  1.42& 0.17&  1.55& 0.23\\
         &$\Delta/\sigma(\Delta)$        &  2.08& 0.22&  2.08& 0.17&  2.39& 0.30\\
\hline
         &$\chi^2$                       & 16.36& 0.89& 18.24& 1.13& 17.68& 1.25\\
$B_{sp}$ &$\Delta_{1}/\sigma(\Delta_{1})$&  1.28& 0.11&  1.26& 0.21&  1.28& 0.15\\
         &$\Delta/\sigma(\Delta)$        &  2.07& 0.18&  2.15& 0.16&  2.12& 0.20\\
\hline
         &$\chi^2$                       & 16.56& 1.41& 15.57& 1.37& 19.07& 1.77\\
$B_{nsp}$&$\Delta_{1}/\sigma(\Delta_{1})$&  1.43& 0.12&  1.38& 0.20&  1.78& 0.26\\
         &$\Delta/\sigma(\Delta)$        &  1.97& 0.15&  1.94& 0.20&  2.19& 0.28\\
\hline
\end{tabular}
\end{table}
 
\end{document}